\documentclass[aps,prl,showpacs,superscriptaddress,twocolumn]{revtex4-1}
\usepackage{amsmath,graphicx}
\usepackage{hyperref}
\begin{document}
%
%%%%%%%%%%%%%%%%%%%%%%%%%%%%%%%%
%            Title             %
%%%%%%%%%%%%%%%%%%%%%%%%%%%%%%%%
%
%
\title{Dynamical arrest of ultracold lattice fermions}
\author{Bernd~Schmidt}
\affiliation{Institut f\"{u}r Theoretische Physik,
Goethe-Universit\"{a}t Frankfurt, 60438 Frankfurt/Main, Germany}
\author{M.~Reza~Bakhtiari}
\affiliation{Institut f\"{u}r Theoretische Physik,
Goethe-Universit\"{a}t Frankfurt, 60438 Frankfurt/Main, Germany}
\author{Irakli~Titvinidze}
\affiliation{Institut f\"{u}r Theoretische Physik,
Goethe-Universit\"{a}t Frankfurt, 60438 Frankfurt/Main, Germany}
\affiliation{I. Institut f\"{u}r Theoretische Physik,
Universit\"{a}t Hamburg, 20355 Hamburg, Germany}
\author{Ulrich~Schneider}
\affiliation{Fakult\"at f\"ur Physik, 
Ludwig-Maximilians-Universit\"at, 80799 M\"unchen, Germany}
\author{Michiel~Snoek}
\affiliation{Institute for Theoretical Physics,
Universiteit van Amsterdam, 1090 GL Amsterdam, The Netherlands}
\author{Walter~Hofstetter}
\affiliation{Institut f\"{u}r Theoretische Physik,
Goethe-Universit\"{a}t Frankfurt, 60438 Frankfurt/Main, Germany}
%
%%%%%%%%%%%%%%%%%%%%%%%%%%%%%%%
%          Abstract           %
%%%%%%%%%%%%%%%%%%%%%%%%%%%%%%%
%
\begin{abstract}
We theoretically investigate the thermodynamics of an interacting
inhomogeneous two-component Fermi gas in an optical lattice. 
Motivated by a recent experiment by L. Hackerm\"uller 
\textit{et al.}, Science, \textbf{327}, 1621 (2010), we study the effect of the 
interplay between thermodynamics and strong correlations on the size
of the fermionic cloud. We use dynamical mean-field theory to compute
the cloud size, which in the experiment shows an anomalous expansion
behavior upon increasing attractive interaction. We confirm this
qualitative effect but, assuming adiabaticity, we find quantitative
agreement only for weak interactions. For strong 
interactions we observe significant non-equilibrium effects which we 
attribute to a dynamical arrest of the particles due to increasing correlations.
\end{abstract}

\maketitle
%
%
%%%%%%%%%%%%%%%%%%%%%%%%%%%%%%%
%        Introduction         %
%%%%%%%%%%%%%%%%%%%%%%%%%%%%%%%
%
%
\textit{Introduction---} 
At the heart of many condensed-matter phenomena lies the interplay 
between strong correlations and temperature. However, even the 
minimal model incorporating these effects, the Hubbard model, 
withstands an exact solution. Ultracold quantum gases in optical 
lattices provide a new way to emulate the physics of these model
systems in a highly controlled way \cite{Bloch2008}: not only can almost all system 
parameters be tuned with ultimate precision, but also the microscopic
details underlying the model are fully known. In this way the phase
diagram of the Bose-Hubbard model, which includes the bosonic Mott
insulator, has been mapped out \cite{Greiner2002}. For fermionic
lattice systems, which directly correspond to electrons in 
crystalline solid-state lattices, the 
formation of an incompressible Mott insulating state has been 
observed as well \cite{Joerdens2008, Schneider2008}.
Using Feshbach resonances, it has also been possible to implement the 
attractive Hubbard model \cite{Strohmaier2010,Hackermuller2010}, which sustains an s-wave superfluid state 
at low temperature and entropy.

One ultimate goal of this research direction is to establish a
detailed description of the Fermi-Hubbard model applicable to the
strongly-correlated regime of high-$T_c$ superconductors
\cite{Esslinger2010,Hofstetter2002}. 
Unfortunately, current experimental
entropies are too high to observe phenomena such as magnetic
ordering \cite{McKay2011}.  
While average entropies per particle down to $S/k_BN\sim 0.5$ have
been demonstrated for fermions in pure dipole traps \cite{Ho2009}, in 
optical lattices only values down to $S/k_BN\sim 1-2$ could be
realized so far \cite{Hackermuller2010,Greif2011},
which is well above the maximum entropy where antiferromagnetic
ordering can be observed \cite{Fuchs2011,Paiva2011}.

However, even at the current experimental entropies, interesting
physics emerges from the interplay between strong correlations and
thermodynamics, as has been studied by several authors for repulsive
interactions \cite{Joerdens2010,DeLeo2008,Fuchs2011,Khatami2011} in the context of Mott insulators. 
Recently, Hackerm\"uller \textit{et al.} experimentally
\cite{Hackermuller2010} investigated this interplay between strong
correlations and entropy, focusing on the attractive regime. 
By loading a two-component Fermi gas into
a three-dimensional (3D) optical lattice in the presence of a
harmonic trapping potential, the size of
the fermionic cloud was measured for different interaction strengths. 
One would naturally expect that this leads to an increasing cloud
size for repulsive interactions while an increasing attractive interaction
should lead to a decreasing cloud size. However, a counterintuitive behavior of the
cloud size was observed: the cloud shrinks upon entering the
attractive regime, but reaches a minimum at relatively small
attractive interaction. For larger attraction the cloud expands
again. 

This anomalous trend was attributed to the adiabatic heating effect: 
for sufficiently strong attractive interaction, singly occupied sites
have a much higher energy than doubly occupied ones and become
energetically irrelevant. This strongly reduces the available Hilbert
space and, at constant temperature, leads to a lower entropy. In
order to keep the entropy constant, the temperature
increases, leading to a higher entropy in the motional degree of
freedom and therefore to the anomalous expansion.

In this Letter we theoretically investigate this experiment by means
of Dynamical Mean-Field Theory (DMFT) simulations of the
Hubbard model in the presence of a harmonic trap. 
Even though our calculations fully include the adiabatic heating effect
and also find an anomalous expansion of the cloud size for large 
attractive interactions, we only find an agreement between our 
equilibrium calculation and the experimental
data of \cite{Hackermuller2010} 
for weak interactions, whereas we find a significant discrepancy for strong interactions.
We attribute this to a strong slowing down of particle transport 
already for moderately large interactions, leading to a \emph{dynamical arrest}
in a state whose size is significantly larger than the equilibrium radius of the particle cloud.
The long-lived arrested states, which we find for lattice
fermions, bear a strong resemblance to those of attractive
colloidal glasses~\cite{Dawson2002}.\looseness=-1 

\textit{Model and Method---} We consider a two-component mixture of
fermionic atoms, loaded into the lowest band of a 3D cubic optical
lattice in the presence of an external harmonic potential. For sufficiently deep optical lattices, this
system is well described by the inhomogeneous Fermi-Hubbard
Hamiltonian
\begin{equation*}
\mathcal{H}\hspace{-.5mm}=\hspace{-.5mm}-J\hspace{-2mm}\sum_{\langle i,j\rangle,\sigma}
\hspace{-1mm}(\hat{c}^\dagger_{i\sigma}\hat{c}^{\phantom{\dagger}}_{j\sigma}
+h.c.)+U\hspace{-1mm}\sum_i \hspace{-1mm}\hat{n}_{i\uparrow}\hat{n}_{i\downarrow}
\hspace{-1mm}-\hspace{-1mm}\sum_{i\sigma}(\mu-V_0r_i^2)\hat{n}_{i\sigma}.\label{Hamil}
\end{equation*}

Here $\hat{c}^\dagger_{i\sigma}$ and 
$\hat{c}_{i\sigma}^{\phantom{\dagger}}$ are the fermionic creation
and annihilation operators and $\hat{n}_{i\sigma}=
\hat{c}^\dagger_{i\sigma}\hat{c}^{\phantom{\dagger}}_{i\sigma}$ is
the number operator, where $\sigma\in\{\uparrow,\downarrow\}$ labels
the two hyperfine states. The on-site interaction is denoted by $U$,
the single-atom hopping amplitude between nearest neighbors
$\langle i,j\rangle$ is $J$  and $\mu$ is the chemical potential
controlling the particle number.
We consider a pancake shaped harmonic trapping potential of strength
 $V_0$ with aspect ratio $\gamma$ and define 
$r_i^2=x_i^2+y_i^2+\gamma^2z_i^2$ as the squared distance of site
$i$ from the trap center. 

\begin{figure}[t]
\includegraphics[width=8.5cm]{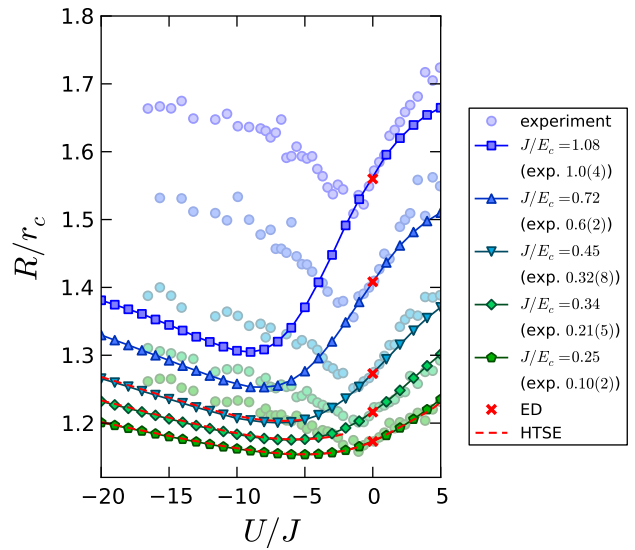}
\caption{\label{fig1}Comparison of experimental cloud radius with
DMFT results \cite{comment_exp}. While we find good agreement for small 
interactions, the experimentally observed radii are significantly 
higher than expected from theory for stronger interactions. We attribute 
this to the dynamical arrest discussed in the main text. 
Also shown are the results of 
a 6th order high temperature series expansion (HTSE) of the Hubbard
model (dashed lines) \cite{Henderson1992,tenHaaf1992}. Since temperature
decreases with increasing $J/E_c$ and $U/J$ (cf. Fig.~\ref{fig2}~(left))
the HTSE is only applicable to sufficiently small or negative $U/J$
and small $J/E_c$.}
\end{figure}

\begin{figure}[t]
\includegraphics[width=8.5cm]{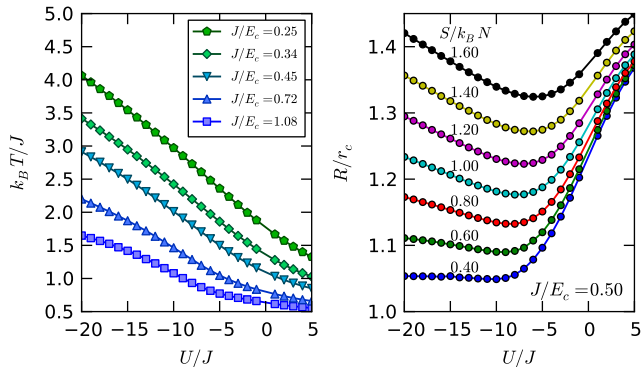}
\caption{ \label{fig2}\textbf{Left}: Temperature vs interaction for an
entropy per particle of $S/k_BN=1.15$ and different external harmonic
confinements. \textbf{Right}: Cloud radius for different entropies
$S/k_BN=0.4-1.6$. The anomalous expansion becomes more pronounced for
higher entropies.
}
\end{figure}

\begin{figure}[t]
\includegraphics[width=8.5cm]{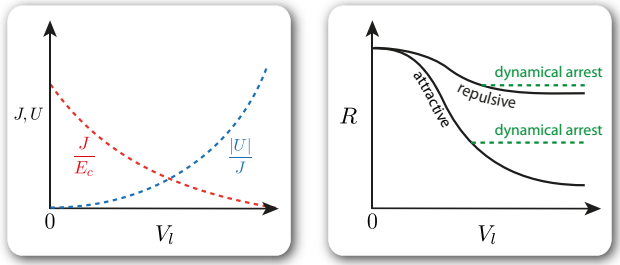}
\caption{\label{cartoon}Interacting fermions during the lattice 
ramp up: In the experiment, the harmonic confinement $E_c$ and the 
scattering length are set in the dipole trap and remain fixed during
the lattice ramp. \textbf{Left}: The tunneling and interaction during the ramp.
\textbf{Right}: Equilibrium cloud size for attractive 
and repulsive interactions during the ramp. The dashed line denotes the non-adiabatic path of the dynamical arrest.}
\end{figure}

\begin{figure}[b]
\includegraphics[width=8.5cm]{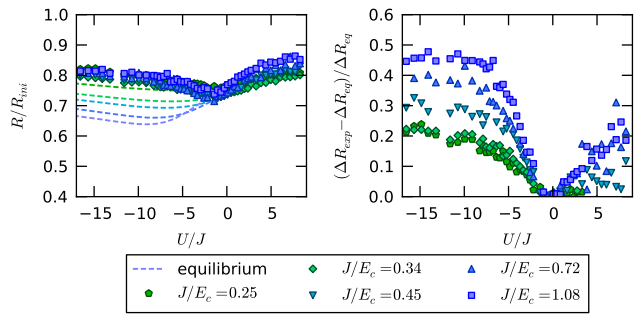}
\caption{\label{fig:shrink}%
\textbf{Left}: Final radius $R$ of the cloud relative to the initial radius $R_\text{ini}$ before ramp-up of lattice.
(symbols: experiment, lines: equilibrium theory, same colors denote same $J/E_c$)
\textbf{Right}: Relative difference between expected ($\Delta R_\text{eq}=R_\text{eq}-R_\text{ini}$) 
and observed ($\Delta R_\text{exp}=R_\text{exp}-R_\text{ini}$) change in cloud size
during the lattice loading from the expected change in cloud size during the ramp. 
}
\end{figure}

\begin{figure}[t]
\includegraphics[width=8.5cm]{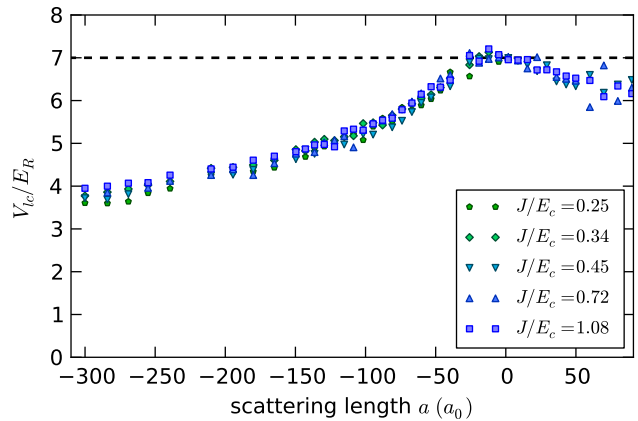}
\caption{\label{fig3} Critical lattice depth calculated from the
experimentally measured radius via Eq. (\ref{eq:radius}), as a
function of the scattering length. The dashed line indicates the final
maximum lattice depth at $V_l/E_R=7$ to which the optical lattice was
ramped up. The values for $J/E_c$ given in the legend are reached for lattice depth $7E_R$.  
}
\end{figure}

We apply DMFT to obtain the \emph{equilibrium} properties of this
Hamiltonian. Employing the assumption of a local self-energy, DMFT
treats local quantum correlations in a fully non-perturbative manner
\cite{Georges1996}. The Fermi-Hubbard model is mapped to an effective
single-site impurity problem coupled to a non-interacting bath.
Here we use exact diagonalization to obtain the self-consistent
solution of the impurity model. To incorporate the harmonic
potential, we use the local density approximation (LDA), which is
known to reliably predict the density distribution for fermionic systems when the trap is
sufficiently shallow \cite{Helmes2008,Snoek2008}. In this approximation, every
site is modeled as part of a homogeneous system with local chemical
potential $\mu(r_i)=\mu_0-V_0r_i^2$. We obtain the density profile
of the fermions, from which the cloud radius is extracted.
To make a direct comparison with the experiment, we employ the following
definition for the cloud radius \cite{Hackermuller2010}
\begin{equation}
 R^2=\frac{1}{N_\sigma}\sum_{i}r_i^2\,\langle \hat n_{i,\sigma}^{\phantom{2}}\rangle,\label{eq:Radius}
\end{equation}
where $N_\sigma$ is the total particle number per spin state. We consider
a balanced mixture with total particle number 
$N=2N_\sigma$, such that $R$ in Eq. (\ref{eq:Radius}) is independent of
$\sigma$. To compute the entropy per lattice site, we use the Maxwell
relation $s(\mu)=\int_{-\infty}^\mu d\mu'\partial\,n(\mu')/\partial T$ and
then obtain the total entropy by $S=\sum_i s(\mu(r_i))$. The harmonic potential is characterized by the
characteristic energy 
$E_c=V_0r_c^2=\frac{3}{5}V_0(3\gamma N/8\pi)^{2/3}$, which is the
mean potential energy per particle of a maximally 
packed state at the bottom of the trap with total particle number
$N$; $r_c$ is the corresponding radius, which is used as the length
scale to express the cloud size $R/r_c$. We note that in LDA the
rescaled radius $R/r_c$ is fully determined by the average entropy
per particle $S/k_BN$ and the ratios of tunneling to trap strength
$J/E_c$ and interaction to tunneling $U/J$, implying that all
results shown in Fig.~\ref{fig1}-\ref{fig3} are independent of
the details of the trap. The experiment has been performed for $N_\sigma \approx  1.5\times 10^5$ particles at a
temperature $T/T_F = 0.12 \pm 0.03$ before loading of the lattice (where $T_F$ is the Fermi temperature) 
and an entropy per particle of $S/k_BN = 1.15 \pm 0.25$. In the experiment, 
both $S/k_BN$ and $J/E_c$ were known only
with a considerable uncertainty \cite{Hackermuller2010}. Since the effect of the experimental
uncertainty in $J/E_c$ on the cloud size is larger, we chose to
perform all calculations for a fixed entropy per particle of
$S/k_BN=1.15$ and determined the value of $J/E_c$ by comparing the experimental radius for
$U/J=0$ with exact diagonalization
(ED) results. In the non-interacting case non-equilibrium effects are
minimal, so that the experimental data at $U/J=0$ can be reliably fitted to 
theoretical calculations \cite{Schneider2012}. The resulting $J/E_c$ agree with 
the experimental data within errorbars for low confinements, and only deviate 
at high confinements, where also some heating was observed in the experiment 
(cf. supporting online material to \cite{Hackermuller2010}). In addition we find satisfying quantitative agreement
with our DMFT calculations for the fixed $J/E_c$ (without additional fitting parameters) in the regime of weak interactions.

%
%%%%%%%%%%%%%%%%%%%%%%%%%%%%%%%
%          Figure 1           %
%%%%%%%%%%%%%%%%%%%%%%%%%%%%%%%
%
%
%
%%%%%%%%%%%%%%%%%%%%%%%%%%%%%%%
%             Results         %
%%%%%%%%%%%%%%%%%%%%%%%%%%%%%%%
%
%
\textit{Results---} 
In Fig.~\ref{fig1} we show our DMFT results for the rescaled cloud
radius $R/r_c$ versus $U/J$ for various
ratios of tunneling to trap strength $J/E_c$, which are compared
directly to the experimental data points \cite{Hackermuller2010}.
Also a comparison with a 6th order high temperature series expansion
is included in this figure, which agrees very well with the DMFT
calculations. While we included the possibility of $s$-wave
superfluid order in our calculations, we found that the experimental
entropy is too high for superfluidity to be present.

For weak interactions $|U/J| \lesssim2$, we
observe very good quantitative agreement with the experimental data
points, evidencing that the DMFT solution of the single-band Hubbard
model incorporates the relevant physics in this parameter regime. 
For stronger attractive interactions we also find the anomalous
expansion of the cloud size in the DMFT calculations. By plotting the
temperature at constant entropy, we can directly verify that this is
indeed due to adiabatic heating originating from a reduced available
phase space. This is displayed in Fig.~\ref{fig2}~(left) and indeed shows
a strong rise for large attraction. When the entropy is decreased,
the anomalous expansion is expected to disappear, since essentially
ground state properties are probed in this case. This is clearly
visible in Fig.~\ref{fig2}~(right), where the radius for different entropies
is shown. For repulsive interactions an adiabatic cooling effect
is observed, similar to what was found in \cite{Paiva2011} and \cite{Schneider2008} (supporting online material).
Although for strong repulsion a similar reduction in phase space
occurs, because doubly occupied sites are energetically unfavored,
this is more than compensated by additional (spin) configurational entropy
for the fermions at the wings of the clouds. 

Although our simulations thus fully include the adiabatic heating effect, 
our results show that the anomalous expansion seen in the experiment 
cannot be explained by adiabatic heating alone:
At stronger interactions we find significant deviations from the
experimentally observed cloud sizes, which are much larger than theoretically expected.
%
%%%%%%%%%%%%%%%%%%%%%%%%%%%%%%%
%          Figure 2           %
%%%%%%%%%%%%%%%%%%%%%%%%%%%%%%%
%
By measuring the temperature after unloading the atoms again from the 
lattice it was excluded that the discrepancy is due to heating \cite{Hackermuller2010}.
We attribute this difference to strong non-equilibrium
effects in the loading: after evaporative cooling, the cloud size in
the pure harmonic trap is significantly larger than the final equilibrium size in the lattice.
During the ramp-up of the lattice, the effective mass of the atoms
increases due to a reduction of kinetic energy in the lowest band
of the lattice. This leads to a shrinking of the cloud (see Fig.~\ref{cartoon}) under adiabatic conditions. 
The scattering length, which
is adjusted by using a Feshbach resonance before the loading of the
lattice, gives rise to an interaction strength $U/J$ which
rapidly grows with the lattice depth. As recently demonstrated
\cite{Schneider2012}, even moderate interactions slow down the atoms
severely, thereby prohibiting the large-scale particle transport necessary for following the adiabatic path.
This can be seen e.g. by comparing the initial radius before 
the ramp-up with the radius in the lattice (Fig.~4~(left)): In 
the experiment, the radius shrinks always by the same percentage
independently of the trap strength, which would not be the case 
if the cloud stays in equilibrium. 
In contrast to the situation in \cite{Schneider2012}, where a
symmetric reduction of particle transport for attractive and repulsive
interactions was observed, we observe much weaker non-adiabatic effects
for the repulsive regime in this case. This is probably due
to the fact that the difference between the radius before the ramp-up
and the equilibrium radius at the final lattice depth is much smaller
in the repulsive than in the attractive regime. Once this effect 
is scaled out (see Fig.~\ref{fig:shrink}~(right)) we see a significant 
deviation on both sides. Moreover, repulsive interactions decrease the 
number of double occupancies while attractive interactions increase it, thereby 
giving rise to different dynamics.
Interestingly, we see that the deviation in Fig.~\ref{fig:shrink} (right) 
becomes independent of $U/J$ for sufficiently strong attraction. This is 
because the experimental and theoretical curves in Fig.~\ref{fig1} - although offset - 
have the same slope.     

To quantify the proposed scenario, i.e. that the disagreement between the experimental and theoretical 
cloud size has to be attributed to a dynamical arrest of the particles due to the slowing down 
of particle transport, we substitute this gradual change in adiabaticity by a step-wise model where up to
a critical lattice depth $V_{lc}$ the system is in thermal and chemical equilibrium and
experiences fully adiabatic changes of its state.  When the lattice depth exceeds a
critical value we assume the cloud to be completely static, as illustrated in Fig.~ 
\ref{cartoon}~(right). 

Within this model we only need
equilibrium  simulations to determine the critical lattice depth. 
As seen before, within LDA the rescaled
radius $\tilde R=R/r_c$ depends only 
on $\tilde S=S/k_BN$,  $\tilde J=J/E_c$ and $\tilde U=U/J$. 
This allows us
to determine the lattice depth at which the system freezes by calculating
$\tilde J = \tilde J(\tilde V_l)$ and $\tilde U =
\tilde U (\tilde V_l,\tilde a)$ as a function of the lattice depth
$\tilde V_l = V_l/E_R$ using band structure
calculations.
Here $\tilde a=a/a_0$ ($a_0$: Bohr radius) denotes the scattering
length. We take into account that $E_c$ also depends on
the lattice depth due to the anticonfining effect of the optical
lattice. The critical lattice depth $V_{lc}$ is then defined as the value
where the calculated equilibrium radius coincides with
the experimentally measured one, i.e.:
\begin{equation}
\tilde R_\text{exp}=\tilde R(\tilde S,\tilde J(\tilde V_{lc}),\tilde U(\tilde V_{lc},\tilde a)),
\label{eq:radius}
\end{equation}
where $\tilde R_\text{exp}$ is the experimentally measured radius
and the function $\tilde R$ is calculated with DMFT.

%
%%%%%%%%%%%%%%%%%%%%%%%%%%%%%%%
%          Figure 3           %
%%%%%%%%%%%%%%%%%%%%%%%%%%%%%%%
%
%
%%%%%%%%%%%%%%%%%%%%%%%%%%%%%%%
%          Figure 4           %
%%%%%%%%%%%%%%%%%%%%%%%%%%%%%%%
%
%
%%%%%%%%%%%%%%%%%%%%%%%%%%%%%%%
%          Figure 5           %
%%%%%%%%%%%%%%%%%%%%%%%%%%%%%%%
%

The resulting critical lattice depths $V_{lc}$ for dynamical
arrest of the cloud are shown in Fig.~\ref{fig3}.
First, we observe a strong dependence on the scattering length,
which is indeed fully consistent with the dynamical arrest hypothesis, 
as the scattering length linearly affects $U/J$, which determines the 
transport properties. We note that for stronger 
interactions the critical lattice depth is rather low, and almost 
leaves the validity regime of the single-band Hubbard model.
Second, we observe that $V_{lc}$ is relatively independent
of the external harmonic confinement $V_0$, such that the data points
for different $J/E_c$ collapse almost to a single curve. 
Indeed, while a tighter trap increases the density and thereby 
slows down the dynamics, it at the same time also decreases the 
distance for the required particle transport. 

%
%%%%%%%%%%%%%%%%%%%%%%%%%%%%%%%
%          Conclusion         %
%%%%%%%%%%%%%%%%%%%%%%%%%%%%%%%
%
\textit{Conclusion---}
We calculated the cloud size of an interacting Fermi gas by means of
DMFT and obtained very good quantitative agreement with experimental
data for weak interactions. Despite qualitative agreement, we
observe a significant discrepancy with
the experiment for strong interactions, which we attribute to non-equilibrium effects caused
by an interaction-induced slowing down of particle transport. The system is 
therefore dynamically arrested at a critical lattice depth, which we 
observe to be almost independent of the harmonic trap.
Our findings have far-reaching consequences for future experiments on
strongly interacting Fermi gases in optical lattices, for
instance with the goal of observing s-wave superfluidity or 
antiferromagnetic order: According to the scenario described above, 
any loading sequence into a deep lattice should be tailored such that it 
minimizes the required density redistribution, otherwise the
system will always freeze in a non-equilibrium arrested
state, rendering the observation of the desired low-temperature properties challenging.

We would like to acknowledge useful discussions with I. Bloch.
This work was supported by the German Science Foundation DFG
(FOR 801 and SFB/TR 49) and the
Netherlands Organization for Scientific Research (NWO).

\end{document}